\documentstyle[11pt]{article}
\begin{document}

%Options for printing 2 pages per 1 sheet (our driver)
%/tr1         transformation
%/cp2         2 pages to 1
%/cx-5.7in    ofset
%/l5.5in      left margin
%/t-0.6in     top margin
%/cf1         page ordering...
%/b1          first page
%/e10         last page

\hyphenation{Vi-ra-so-ro}

\begin{center}

\title{Two-parametric zeta function regularization in superstring
theory}
\author{Lubo\v s Motl}
\date{October 5th, 1995}
\maketitle

\vspace{4mm}

%\renewcommand
%\baselinestretch{0.8}
%\medskip

{\it
Faculty of Mathematics and Physics\footnote{e-mail:
{\tt motl@port.troja.mff.cuni.cz},

http://www.ms.mff.cuni.cz/acad/webik/~lmot2220/home.html},
Charles University, \\
Prague, Czech republic}

%\renewcommand
%\baselinestretch{1.4}

\vspace{1.5cm}

{\bf Abstract}

\end{center}

\def\ket#1{|#1\rangle}
\def\no#1{:\!#1\!:}            %normal ordering
\def\exc{\leftrightarrow}
\def\eq#1{\begin{equation}#1\end{equation}}
\def\eqa#1{\begin{eqnarray}#1\end{eqnarray}}
\def\zpet{\!\!\!\!\!\!\!\!}

In this paper some quite simple examples of applications of
the zeta-function regularization to superstring theories are
presented.
It is shown that the Virasoro anomaly in the BRST formulation
of (super)strings can be directly computed from the original
expressions of the operators as well as normal ordering constants
and masses of ground levels. Hawking's zeta regularization is
recognized as an efficient tool for direct calculations, bringing
no ambiguities.

Possible implications for global GSO operators' phases
definitions (maybe ensuring modular invariance) will be discussed
elsewhere.

\vspace{1cm}

\begin{center}
\Large{This paper is dedicated to my friend Lenka.}\\
\Large{arch-ive/9510105}
\end{center}

\newpage %end of abstract

\section{Simple examples, usual Riemann zeta function}

I begin with the simplest determination of the bosonic string
critical dimension. In the light-cone gauge the operator $p^-$
\eq{p^-=\frac1{2\pi p^+}\int_0^\pi d\sigma\cdot(\pi^2
p(\sigma)^2+x'^2)}
has form (after translating to modes $\alpha_n$)
\eq{p^-=\frac1{2 p^+}({p^i}^2+m^2)}
where we used (now we consider open strings)
\eq{{p^i}^2+m^2=\sum_{n\in Z}\alpha^i_{n}\alpha^i_{-n}.}
Here $\alpha_n$ are annihilation operators for $n>0$, so if we want
to write
the expression in a normal-ordered form, we must change the order
of the operators for $n>0$ terms, according to the commutation
relation
\eq{[\alpha^i_m,\alpha^j_n]=m^1\delta_{m+n}\delta_{ij}.}
Then we get ($\alpha^i_0=p^i$)
\eq{m^2=2\sum_{n=1}^\infty (\alpha^i_{-n}\alpha^i_n+(d-
2)\frac{n^1}2),}
where $(d-2)=\delta^{ii}$ means number of transversal coordinates.
The first part gives a finite contribution when operating on a state,
particularly the first term annihilates the ground level (tachyon).
The second part is a clearly divergent sum which must be regularized
(see next section). Analytic continuation according to the exponent
$1$ in $n^1$ gives
\eq{\sum_{n=1}^\infty
n^1=1+2+3+4+\dots=0+1+2+3+\dots=\zeta(-1)=-\frac{1}{12}.}
I hope that the reader will not be disturbed too much by the equations
between finite numbers and divergent sums.
(To be provocative, they really equal.)

The first excited level of open string $\alpha^i_{-1}\ket0$ has only
$(d-2)$ degeneracy, so it can't form a massive representation of the
subgroup $SO(d-1)$ of the full Lorentz group, fixing some
$d-$momentum vector of this level. It means that the first level
(for which $\sum \alpha_{-n}^i\alpha^i_n=1$) must be massless.
\eq{m^2=2\left(\sum_{n=1}^\infty(\alpha^i_{-n}\alpha^i_n)-
\frac{d-2}{24}\right)}
Condition $m^2=0$ gives $d=26$.

\subsection{A quick analytic continuation}

Let's consider a bit generalized Riemann zeta function with a
parameter
$s$
\eq{\zeta_s(x)=\sum_{n=1}^\infty (n+s)^{(-x)},}
where the most usual value of $s$ will be $s=0$. This formula is
convergent for $Re\,s>1$, for instance $\zeta(2)=\pi^2/6$ (exactly).
Let's note that if we add $1$ to $s$, the only difference will be
that the first term drops ($n=1$). We can use Taylor series according
to the parameter $s$.
\eq{\zeta_{s+1}(x)=\zeta_s(x)-(1+s)^{-x}=\zeta_s(x)+\frac1{1!}
\frac{\partial \zeta_{s'}(x)}{\partial s'}|_{s'=s}+
\frac1{2!}\frac{\partial^2 \zeta_{s'}(x)}{\partial s'}|_{s'=s}+\dots}
But the derivatives with respect to $s$ are computed easily:
\eq{\frac{\partial}{\partial s}\zeta_s(x)=\sum_{n=1}^\infty (n+s)^{-
x}=(-x)
\zeta_s(x+1)}
and $m$-th derivative
\eq{\frac{\partial^m}{\partial s^m}=(-x)(-x-1)\dots(-x-
m+1)\zeta_s(x+m)}
So we have
\begin{eqnarray}
-(1+s)^{-x}=(-x)\zeta_s(x+1)+\frac{(-x)(-x-1)}{2!}\zeta_s(x+2)+\\
+\frac{(-x)(-x-1)(-x-2)}{3!}\zeta_s(x+3)+\nonumber\dots
\end{eqnarray}
Let's substitute $x\to 0$. Since for $x>1$ zeta function is finite
and the terms are multiplied by $(-x)\to 0$, on the right hand side
just the first term survives, e.g.
\eq{\lim_{x\to 0} x\zeta_s(x+1)=1}
$\zeta_s(y)$ has pole for $y\to1$.
Substituing $x\to -1$ gives
$$-1+s=\zeta_s(0)+\frac{1}{2!}(-x-1)\zeta_s(x+2)$$
But $\lim_{x\to-1}(-x-1)\zeta_s(x+2)=-1$, therefore
\eq{\zeta_s(0)=-\frac12-s}
Substituing $x\to -2$ gives equation
$$-(1+s)^2=2\zeta_0(-1)+\frac{1}{2!}2\cdot 1
\zeta_s(0)+\frac{1}{3!}2\cdot1\cdot
(-x-2)\zeta_0(x+3)$$
which after short algebra
\eqa{-(1+s)^2=2\zeta_s(-1)+(-\frac12-s)+(-\frac13),\nonumber\\
\zeta_s(-1)=-\frac{1}{12}-\frac{s+s^2}2=\frac1{24}-\frac{(s+1/2)^2}2}
Without details we mention also $(x\to -3)$
\eq{\zeta_s(-2)=-\frac{s(s+1/2)(s+1)}3.}
Other values we give for $s=0$ only
\eq{\zeta_0(-1)=-\frac{1}{12},\,\,\zeta_0(-3)=\frac1{120},\,\,
\zeta_0(-5)=-\frac1{252}}
Interesting fact that zeta of negative even number vanishes
\eq{\zeta_0(-2)=\zeta_0(-4)=\zeta_0(-6)=\dots=0}
can be proved by mathematical induction if we sum
the Taylor series around $s=0$ for $s=+1$ and $s=-1$ but I will
not enter to details.

\subsection{Consistency of the regularization}

Presented formulas for the divergent sums have many characteristics
of consistency. For example, if we make $s$ increase by one, the
result decreases
by the first term (this fact was used in the derivation).

For example (for other cases similarly)
\eq{1+\sum_{n=y+1}^\infty n^0=1+\frac 12-(y+1)=\sum_{n=y}^\infty n^0.}

So we regularize in fact really only the part ``in far infinity''
and with any finite number of terms we can manage as with normal
numbers.

Next interesting property states that
\eq{\sum_{n\in Z} (n+s)^x=0\quad\mbox{for $x=0,1,2,$\dots}.}
This identity is quite important in checking the independence
of the commutators on a particular form of the commutants.
Particularly, $\sum_{n=1}^\infty n^0=\sum_{n=-1}^{-\infty}n^0=-1/2$
and these $-1/2$'s together with $1$ arising from $n=0$ give zero.

%\subsection{Correspondence with lattice-cutoff regularization}

\section{Modifications of formulas for the regularization}

In the first example we noted that the regularization parameter is
the exponent over the mode's index. It means that in a sense we
compute
the result of a regularized parameter for a general (complex) degree
of the derivatives in this expression, then we realize that the result
is an analytic function of these parameters that can be continued to
the interesting values for the degrees.

Particularly, we must modify some formula for (anti)commutators and
so on.
So for instance, instead of $\{c_m,b_n\}=\delta_{m+n}$ we write a more
precise equation
\eq{\{c_m,b_n\}=\delta_{m+n}\cdot n^0,\quad
\{c_m,c_n\}=\{b_m,b_n\}=0.}
Here we must understand why there is no contradiction in computing
sum $\sum_{n=1}^\infty n^0=1+1+1+\dots=-1/2$. Someone could find it
counterintuitive since if we add $1$ to this sum, we get
some $1+1+1+\dots$ sum again. But the right hand side should equal
$+1/2$. In fact, there is no contradiction here since we must
always remember from which mode the number $1$ arises. In other words,
we should always write it as $n^0$.

\section{Virasoro anomaly}

Different contributions to the (super)Vi\-ra\-so\-ro algebra's anomaly
can be evaluated by worldsheet methods, but we can also use
calculations involving modes. So for example, in the page
I/130 of [GSW], authors argue that the easiest and safest
way to determine the anomaly is by evaluating specific matrix
elements.
But here I wish present a proper way for its direct calculation.

\subsection{Ghost contribution}

In this subsection $L_m$ will always denote $L_m^{(gh)}$. Parameter
$J$
is the conformal dimension of the antighost and $J=2$ for the ordinary
antighost $b$. We define the ghost contributions to Virasoro algebra
generators as
\eq{L_m=\sum_{n\in Z}\left(m(J-1)-n\right)b_{m+n}c_{-n}.}
This expression equals its normal-ordered for $m\neq 0$. For $m=0$ it
differs
from its normal-ordered part by a $c$-number, which we again compute
by zeta-function regularization. (Let's remark
that in [GSW] they always mean $\no{L_m}$ when they use symbol $L_m$.)
Products of ghost operators must be exchanged for $n>0$ (or $n\geq0$
which gives the same result) in the following sum
and a normal ordering constant appears.
\eq{L_0=\sum_{n\in Z} (-n)b_nc_{-n}=\no{L_0}+\sum_{n=1}^\infty(-
n)\cdot n^0=
\no{L_0}+\frac{1}{12}}
Our main task is to compute the commutators of two $L_m$'s.
\eqa{
[L_m,L_{m'}]=\sum_{n,n'\in Z}(m(J-1)-n)(m'(J-1)-n')\cdot\nonumber\\
\cdot\left(b_{m+n}\{c_{-n},b_{m'+n'}\}c_{-n'}-b_{m'+n'}\{c_{-n'},
b_{m+n}\}c_{-n}\right)}
The bottom line contains two terms such that the second can be
obtained from the first one by $(m,n\exc m',n')$. We use the
anticommutators
from the previous section. Kronecker's delta will reduce the
summation over $n,n'$ to only one summation.
$$[L_m,L_{m'}]=\dots=\sum_{n\in Z}(m(J-1)-n)(m'J-n)b_{m+n}c_{m'-
n}\cdot n^0 -
(m,m'\,\exc\,n,n')$$
The ghost oscillators $c,b$ in the last expression can be exchanged
for
$m+m'\neq 0$ since their anticommutator equals zero. Then the result
contains only a finite number of terms with creation operators
$c_{k<0},b_{k<0}$
as the last factors in the products. Therefore $n^0$ can be replaced
by $1$
and terms can be summed classically (no divergent sum need to be
regularized) and we get
\eq{[L_m,L_{m'}]=\dots=(m-m')L_{m+m'}+A(m)\delta_{m+m'}.}
For $m+m'=0$ we must compute also the anomaly term $A(m)$.
If $m'=-m$
\eq{[L_m,L_{-m}]=\sum_{n \in Z}(m(J-1)-n)(-mJ-n)b_{m+n}c_{-m-n}\cdot
n^0
-(m\exc -m)=}
Using substitution $n=N-m$ e.g. $N=n+m$ we get
$$=\sum_{N\in Z}(mJ-N)(m(1-J)-N)b_Nc_{-N}(N-m)^0-(m\exc -m)=$$
The terms with $N>0$ should be rewritten in a normal-ordered fashion.
The $q$-number part combines with $N\leq 0$ terms giving $2m\no{L_0}$
and the $c$-number part appears
\eqa{=2m\no{L_0}+\left(\sum_{N=1}^\infty m^2J(1-J)N^0(N-m)^0-
\right.\nonumber\\
\left.-\sum_{N=1}^\infty mN^1(N-m)^0+ \sum_{N=1}^\infty N^2(N-m)^0-
(m\exc -m)\right)}
We involve a new variable $J=(1+k)/2$ and use results
of double-parametric zeta regularization.
$$=2m\no{L_0}+\left(\frac{m^2}4(1-k^2)(-\frac12+\frac m2)
+\frac m{12}-\frac{m^3}4+\frac{m^3}6-(m\exc -m)
\right)$$

The final result reads
\eq{[L_m, L_{-m}]=2m\no{L_0}+\frac{1}{12}(1-3k^2)m^3+\frac m6=2mL_0+
\frac{1}{12}(1-3k^2)m^3.}
We can notice that for $L_m$ we get a simpler expression containing
$m^3$ term only than if we use $\no{L_0}$ (without the natural
normal ordering constant). If we translate $L_m$ to $T_{++}(\sigma)$,
then
only the $\delta'''(\sigma-\sigma')$ anomaly appears.

\subsection{Bosonic coordinates' contribution}

In this subsection $L_m$ will denote $L^{(x)}_m$. Let's accept
\eq{L_m=\frac 12\sum_{n\in Z}\alpha_{m-n}\alpha_n,\quad
[\alpha_m,\alpha_n]
=m^1\delta_{m+n}.}
Then $L_0$ contains again a natural normal-ordering constant.
\eq{L_0=\no{L_0}+\frac 12\sum_{n=1}^\infty n^1=\no{L_0}-\frac 1{24}.}
Since there are 26 coordinates (26 equal contributions to this
constant),
the total
normal-ordering constant is (using last subsection)
\eq{26L_0^{(x)}+L_0^{(gh)}=\no{26L_0^{(x)}+L_0^{(gh)}}-
\frac{26}{24}+\frac1{12}=
\no{26L_0^{(x)}+L_0^{(gh)}}-1.}
Evaluation of the commutator looks like
\eqa{[L_m,L_{m'}]=\frac 14\sum_{n,n'\in Z}
\left(\alpha_{m-n}[\alpha_n,\alpha_{m'-n'}]\alpha_{n'}
+[\alpha_{m-n},\alpha_{m'-n'}]\alpha_n\alpha_{n'}+\right.\nonumber\\
\left.+\alpha_{m'-n'}\alpha_{m-n}[\alpha_n,\alpha_{n'}]+
\alpha_{m'-n'}[\alpha_{m-n}\alpha_{n'}]\alpha_n\right)=
}
For $m+m\neq 0$ the result again equals its normal-ordered part, so
factors
like $(n'-m')^1$ can be replaced by $(n'-m')$ and summed together.
So the result has general form
\eq{[L_m,L_{m'}]=(m-m')L_{m+m'}+A(m)\delta_{m+m'}.}
The anomaly can be computed using
\eq{\sum_{N=1}^\infty (N+m)^1N^1=\zeta_{0,m}(-1,-1)=\frac{m^3-m}{12}}
Finite anomaly equals
\eq{[L_m,L_{m'}]=2m\no{L_0}+\frac{m^3-m}{12}=2m L_0+\frac{m^3}{12}}
The cancellation with ghost contribution now gives the
critical dimension in form $(k=3);\,d=3k^2-1=26$.

Anomalies in other commutators and anticommutators can be
computed in similar fashion.

\subsection{Two-parametric zeta function}

During the calculations we often needed not only sums like $\sum_n
n^x$
but also a more general
\eq{\zeta_{s,t}(x,y)=\sum_{n=1}^\infty (n+s)^{-x}(n+s+t)^{-y}}
This reduces to previous zeta functions for $t=0$
\eq{\zeta_{s,0}(x,y)=\zeta_s(x+y)}
Also, the sum should be independent on ordering of the product inside
the sum,
e.g.
\eq{\zeta_{s+t,-t}(y,x)=\zeta_{s,t}(x,y)}
The sums for general (nonzero) $t$ can be effectively computed
using Taylor expansion according to $t$.

Direct substitutions gives for $x+y\to0$
\eq{\zeta_{s,t}(1+x,y)=\frac1{x+y}+\mbox{finite part}.}
Calculation of e.g. $\zeta_{s,t}(-m,0)$ runs as follows
(we write the same $\varepsilon$ to both parameters which turns
to be the simplest way to obey equation (38) -- more
precise ways bring the same result)
\eq{\zeta_{s,t}(-m+\varepsilon,\varepsilon)=\zeta_{s,0}(-m,0)+
(-\varepsilon)t\zeta_{s,0}(-m,1)+\dots}
Here only the first term $\propto t^0$ of the Taylor sum and the term
proportional to $\zeta(1+x,y)$, $x+y\to 0$ contribute, giving
\eq{\zeta_{s,t}(-m,0)=\zeta_s(-m)-\frac{t^{m+1}(-1)^m}{2(m+1)}.}
Other values can be deduced in a similar manner.

Here I summarize some useful formulas.
\eqa{
&&\zpet\sum_{n=y}^\infty n^0=\frac 12-y,\quad
\sum_{n=y}^\infty n^1=-\frac 1{12}+\frac{y-y^2}2,\quad
\sum_{n=y}^\infty n^2=-\frac 13y(y-\frac 12)(y-1)\nonumber\\
&&\zpet\sum_{n=y}^\infty n^0(n+m)^0=\frac 12-y-\frac m2,\quad
\sum_{n=y}^\infty n^1(n+m)^0=-\frac 1{12}+\frac{y-
y^2}2+\frac{m^2}4\nonumber\\
&&\zpet\sum_{n=y}^\infty (n+m)^1 n^0=-\frac1{12}+\frac{y-y^2}2+\frac
m2(1-2y-
\frac m2)\\
&&\zpet\sum_{n=y}^\infty n^1(n+m)^1=-\frac13 y(y-1)(y-\frac 12+\frac
32m)+
\frac{m^3-m}{12}\nonumber\\
&&\zpet\sum_{n=y}^\infty (n-\frac m2)^1
(n+\frac m2)^1=-\frac 13 y(y-1)(y-\frac 12)+
\frac {m^2}4(y-\frac12)\nonumber}

\section{Anticommutator of worldsheet SUSY currents in 4F models}

In this section I show rather surprising fact in 4F models (FFFF
models
means models in the four-dimensional free fermionic formulation) that
the anticommutator of the worldsheet SUSY current with itself
gives a correct result, the energy-momentum tensor
containg derivatives of the fermion fields. The zeta-function
regularization is being used. The origin of the derivatives in the
result
is again similar to the emergency of anomalies and normal-ordering
shifts of ground levels.

We repeat the supercurrent in the case of six compactified
dimensions from [af].
\eq{T_F(z)=\psi^\mu\partial_zX_\mu+\sum_{i=1}^6 \chi^i y^i
\omega^i,}
where $\mu=0,1,2,3$, $\psi_\mu$ are the worldsheet superpartners of
bosonic coordinates $X_\mu$ and $\chi,y,\omega$ are also
hermitean fermionic fields.

Now the anticommutators of two $T$ should give an observable
proportional to energy-momentum. ($J_+=T_F$.)
\eq{\{J_+(\sigma),J_+(\sigma')\}\propto\delta(\sigma-\sigma')
T_{++}(\sigma).}
Anticommutator of the first four terms $\psi^\mu\partial_zX_\mu$
gives an expected result. But it seems hard at first look to
obtain terms like $i\cdot y^i\partial_\sigma y^i$ (and also for
the fields $\omega, \chi$) from anticommutator of
terms $\chi y \omega$ containing no derivatives, which have
anticommutation relations as
\eq{\{\chi(\sigma),\chi(\sigma')\}=
\{y(\sigma),y(\sigma')\}=
\{\omega(\sigma),\omega(\sigma')\}\propto \delta(\sigma-\sigma').}
But these derivative terms are obtained due to the similar
phenomenon which causes also the normal-ordering constants,
anomalies in (super)Vi\-ra\-so\-ro algebra, term proportional to $c_0$
in the BRST operator and other\dots

The calculation will be only sketched here and overall
normalization will be ignored. Writing $y(\sigma)$ in terms of
modes $\sum_{n\in Z}e^{2in\sigma}y_n$ (and identical sums for
$\chi$ and $\omega$), and also $J_+=\sum_n F_ne^{2in\sigma}$
we get
\eq{\{F_m,F_{m'}\}\propto \sum_{
 n_1+n_2+n_3=m\over n'_1+n'_2+n'_3=m'
}
y_{n_2}\omega_{n_3}
y_{n'_2}\omega_{n'_3}\delta_{n_1+n'_1}(n_1)^0+
\mbox{($\chi,y,\omega$ cycl.perm.)}}
Cyclic permutations can be rewritten in a similar way as the
first term.
\eq{\sum_{n_2,n'_2,n_3}{y_{n_2}}y_{n'_2}\omega_{n_3}
\omega_{m+m'-n_2-n'_2-n_3}(m-n_2-n_3)^0}
Such an expression can be expressed in the normal-ordered form.
No product of four operators can survive since for such
terms the zeroth power can be omitted and $(n_2\leftrightarrow
n'_2)$ exchange ensures the cancellation.

Only terms with two operators remain. Exchange of the two $y$'s
can contribute by a $c$-number which keeps only $\omega\omega$
terms and vice versa. (Also a total $c$-number anomaly in the
total anticommutator survives but I will not enter to details
here.)

Let us have a look to the $\omega\omega$ terms. They arise from
$y_{n_2}y_{n'_2}$ for $n_2+n'_2=0$, $n'_2\leq0$. So the result
looks something like
\eq{\sum_{n_2=0'}^\infty\sum_{n_3} (n_2)^0 \omega_{n_3}\omega_{m
+m'-n_3}(m-n_2-n_3)^0.}
Symbol $0'$ expresses that only half of the $n_2=0$ term is
summed.
Since $\sum_{n_2=0'}^\infty (n_2)^0 (n_2+n_3-m)^0=1/2-n_3/2+
m/2-1/2=m/2-n_3/2$, we obtain
\eq{\frac 12\sum_{n_3}(m-n_3)\omega_{n_3}\omega_{m+m'-n_3}.}
Because of anticommuting the term $\propto m$ vanishes
(or gives only $c$-number) and the result can be written
also as a mode of $i\omega\partial_\sigma\omega$
\eq{\propto \sum_{n_3}n_3\omega_{n_3} \omega_{m+m'-n_3}.}

\subsection{Other constants in 4F models}

Here I mention that the equation (2) of [af]
\eq{M_L^2=-\frac 12+\frac{\alpha_L\cdot\alpha_L}8+N_L=-1+\frac
{\alpha_R\cdot\alpha_R}8+N_R=M_R^2}
has a natural explanation after using formula
\eq{\sum_{n=y}^\infty n^1=\frac{1}{24}- \frac 12(y-\frac 12)^2.}
Also expression (4) of [af] for the $U(1)$ generator
\eq{Q(f)=\frac12\alpha(f)+F(f)}
can be deduced from the continual definition $\propto
\int_0^\pi d\sigma ff^*$, which after prescription to modes
gives
\eq{\sum_{n\in Z+\frac{\alpha(f)+1}2}  f_{-n}f^*_n=\dots}
(since frequencies of creation operators $f_{-n}$ are
in $Z+(\alpha(f)+1)/2$, e.g. $\alpha=0$ is antiperiodic boundary
condition and $\alpha=1$ periodic), which can be translated
to normal-ordered form by the usual changing order in terms
$n<0$
\eq{\dots=\sum_{n\in Z+\frac{(1-\alpha(f))}{2}} f_nf^*_{-n}=
\no{\mbox{that}}+\frac 12 \alpha(f)=F(f)+\frac 12\alpha(f).}

\subsection{Global GSO operators' phases}

It seems possible that even GSO operators may be defined globally
for all the sectors. Different choices of signs and phases of
GSO operators in different sectors correspond to different
forms how can be these operators written. (They create only
finite numbers of non-equivalent theories.) For example, by
multiplication the operator by
\eq{\mbox{exp}(2\pi i\sum f_nf^*_{-n})}
we change its sign only in periodic sector ($\sum_{n=1}^\infty
n^0=-1/2$) while in antiperiodic sector it remains constant
($\sum_{n=1/2}^\infty n^0=0$).

Models with many sectors can be constructed by defining a group
of unitary GSO operators $\Xi$. Physical state is an eigenstate of all
these operators corresponding to eigenvalue 1.
(This condition remains also for nontrivial sectors.)
So in a sense we ``set the operators equal one'' -- therefore
we must involve for each GSO operator $G\in\Xi$
a sector where a identical operation (particularly
rigid shift of a closed string by $\pi$) has the same effect
as the operator itself. ($G_{NS}$ is the operator anticommuting
with all the fermionic fields, to ensure that the trivial sector
of group $\Xi$ is the {\it antiperiodic} one.)
\eq{G'\cdot L\cdot G'^{-1}=L_{\mbox{\small after\,\,}\sigma
\to\sigma+\pi}, \quad G'=G\cdot G_{NS}}

One of the simplest examples of this process is compactification
on circle. Here the group of sectors (group of GSO operators
$\Xi$) is isomorphic to $Z$. GSO operators are given by ($n\in Z$)
rigid shift of string by multiple of vector $a$
\eq{\mbox{exp}(ina\cdot p_{\mbox{\small zero}}).}
We must include sectors with non-zero winding numbers (where
the shift $\sigma\to\sigma+\pi$ moves the string by $na$).

More conventional example of GSO operator is that change phases
of some of fermionic fields. (Here $g_b$ means a global
phase correction of the operator $G_b$ given by some vector $b$.)
\eq{G_b=g_b\,\mbox{exp}\sum_f b(f)\kappa\int ff^* d\sigma}
where $\kappa$ is a conventional constant such that
$\kappa\int ff^* d\sigma=\sum f_nf^*_{-n}$.

\section{Summary}

Zeta function regularization (given by analytical continuation
of the expressions according to the degrees of derivatives)
was shown as a reliable method
giving correct results in many cases and author argued for using
similar operations with divergent sums as the most direct way
to do the calculations.

%\vspace{3cm}

%\noindent PACS numbers: 03.70.+k, 04.20.Cv, 11.10.Gh

%\newpage

\end{document}